\documentclass[prl,superscriptaddress,twocolumn,showpacs,floatfix,amsmath,amssymb]{revtex4-1}
\usepackage{xcolor}
\usepackage[utf8]{inputenc}
\usepackage{amsmath}
\usepackage{amssymb}
\usepackage{mathtools}
\usepackage{siunitx}
\DeclareSIUnit\bar{bar}
\DeclareSIUnit\ryd{Ry}
\usepackage[version=4]{mhchem}
\usepackage{graphicx}
\usepackage{booktabs}
\usepackage{svg}
\usepackage{gensymb}
\usepackage{graphics}
\usepackage[T1]{fontenc}
\usepackage{soul}

\begin{document}
\title{First-order rhombohedral to cubic phase transition in photoexcited GeTe}

\author{Matteo Furci}
\affiliation{Department of Physics, University of Trento, Via Sommarive 14, 38123 Povo, Italy}
\email{matteo.furci@unitn.it}
\author{Giovanni Marini}
\affiliation{Graphene Labs, Fondazione Istituto Italiano di Tecnologia, Via Morego, I-16163 Genova, Italy}
\author{Matteo Calandra}
\affiliation{Department of Physics, University of Trento, Via Sommarive 14, 38123 Povo, Italy}
\affiliation{Sorbonne Universit\'e, CNRS, Institut des Nanosciences de Paris, UMR7588, F-75252 Paris, France}
\affiliation{Graphene Labs, Fondazione Istituto Italiano di Tecnologia, Via Morego, I-16163 Genova, Italy}
\email{m.calandrabuonaura@unitn.it}

%\email{matteo.furci@unitn.it}

%\affiliation{Department of Physics, University of Trento, Via Sommarive 14, 38123, Povo, Italy}

%\email{giovanni.marini@iit.it}

%\affiliation{Department of Physics, University of Trento, Via Sommarive 14, 38123, Povo, Italy}

%\email{m.calandrabuonaura@unitn.it}

\begin{abstract}
Photoexcited GeTe undergoes a non-thermal phase transition from a rhombohedral to a rocksalt crystalline phase. The microscopic mechanism and the nature of the transition are unclear. By using constrained density functional  perturbation theory and by accounting for  quantum anharmonicity within the stochastic self-consistent harmonic approximation, we show that the non-thermal phase transition is strongly first order and does not involve phonon softening, at odd with the thermal one. The transition is driven by the  closure of the single particle gap in the photoexcited rhombohedral phase. Finally, we show that ultrafast X-ray diffraction data are consistent with a coexistence of the two phases, as expected in a first order transition. Our results are relevant for the understanding of phase transitions and bonding in phase change materials.
\end{abstract}
\maketitle
Phase change materials (PCMs) have found commercial applications in all-optical memories and are promising
candidates for the use in computation beyond Von Neumann \cite{Zhang2019}. Their success is due to two fundamental properties:  i) their amorphous and crystalline phases have a significantly different electrical resistance \cite{ovshinsky1968} and optical properties \cite{lencer2008, Wuttig2017}, and ii) they can be rapidly switched between both states. The electronic structure and structural properties of these materials result from a competition of resonant bonding and Peierls distortion \cite{shportko2008}. 
%\textcolor{red}{The delicate balance between these two phenomena leads to emergence of ferroelectricity at low temperature}. 
PCM structural properties are very sensitive to ultrafast photoexcitation that, in these systems, can trigger phase transitions \cite{matsubara2016}. However, a sound description of the mechanism responsible for the photoexcited phase transitions is lacking even for GeTe, the most simple PCM.

\begin{figure}[b!]
\centering
\includegraphics[width=1\linewidth, trim={0cm 0.5cm .3cm .3cm},clip]{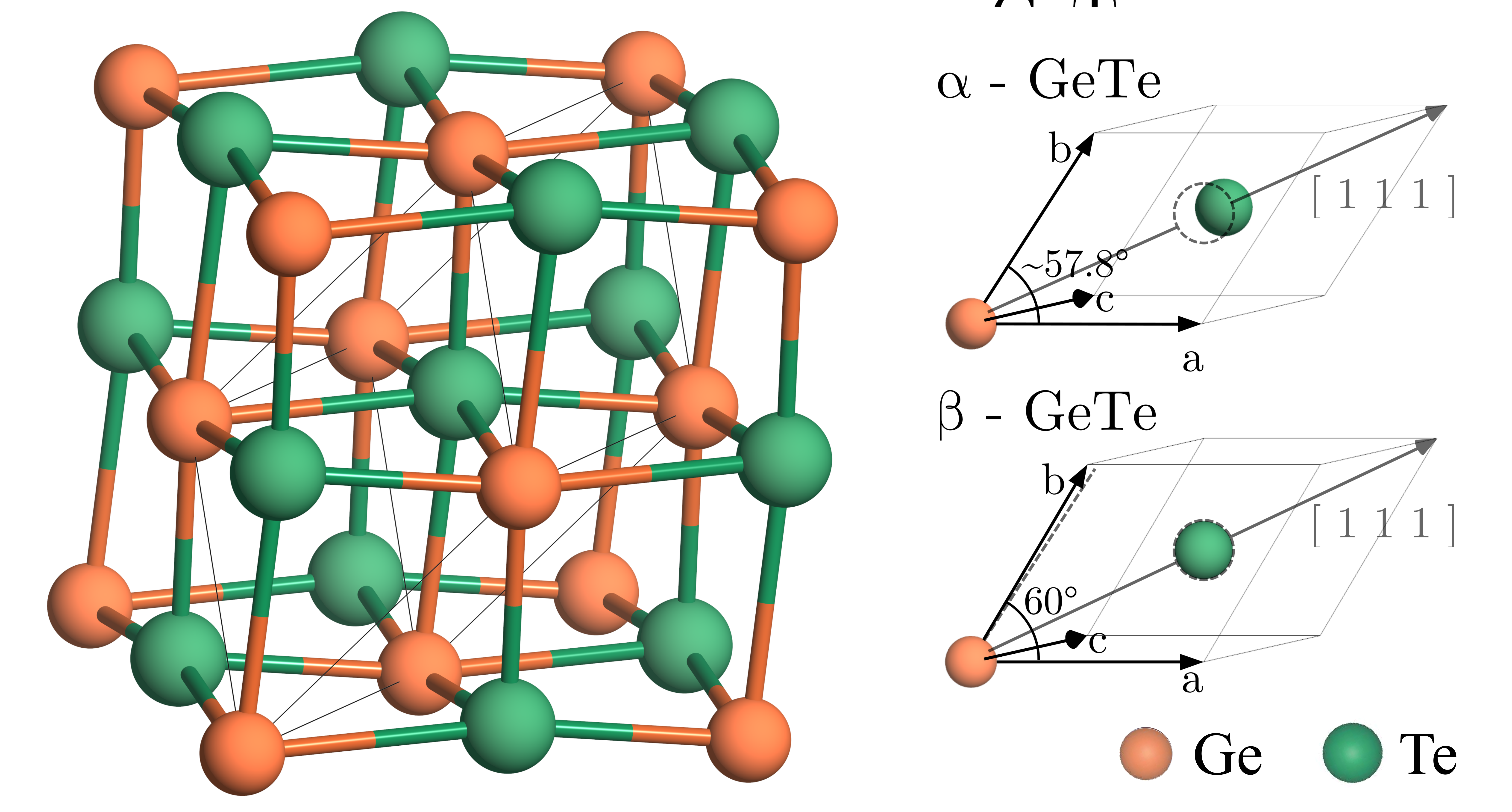}
\caption{Left Panel: pseudo-cubic unit cell of rhombohedral GeTe. The rhombohedral primitive unit cell is shown in black within the pseudo-cubic structure. Right Panel: details of the rhombohedral unit cell for both rhombohedral \ce{GeTe} ($\alpha$-\ce{GeTe}) and cubic \ce{GeTe} ($\beta$-\ce{GeTe}); in both structures, the body diagonal, the geometric centre of the primitive cell and the rhombohedral angle are plotted in gray in order to highlight the differences between the two crystalline phases \cite{Momma:db5098}.}
\label{fig1}
\end{figure}

At ambient conditions, \ce{GeTe} crystallizes in a rhombohedral  phase ($\alpha$-\ce{GeTe}) which exhibits a ferroelectric distortion along the body diagonal of the rhombohedral cell, making $\alpha$-\ce{GeTe} a ferroelectric compound \cite{boschker2017}. For $T \gtrsim \SI{700}{\kelvin}$ \cite{chattopadhyay1987}, \ce{GeTe} adopts a paraelectric rock-salt  crystal structure ($\beta$-\ce{GeTe}) \cite{chatterji2015} (see Fig. \ref{fig1}). 
The thermal displacive phase transition involves an optical phonon softening at zone center, as confirmed by  Raman data \cite{stegmeier1970} and molecular dynamics (MD) simulations \cite{Sosso2012, DangicPhysRevB.106.134113}.
The transition has a  weakly first order character, as shown by the detection of a latent heat \cite{sist2018}.
EXAFS data \cite{fons2010} show an apparent survival of long and short \ce{Ge}-\ce{Te} bonds at $T\ge 700$ K, suggesting an order-disorder phase transition \cite{fons2010}. However,  the latter result has been reinterpreted as due to anharmonicity in the potential in Ref. \cite{DangicPhysRevB.106.134113}. 
The common feature among all these interpretations of the thermal transition is the occurrence of a softening of the zone-center optical phonon mode.

%seem to indicate that, in the thermal case, a first order structural phase transition slightly precedes the  phonon-softening of the displacive transition.
%\textcolor{red}{The nature of the thermal phase transition is debated:}
%Raman data \cite{stegmeier1970} show a softening of the optical modes at zone centre in $\alpha$-\ce{GeTe} pointing to a second-order displacive  transformation, in agreement with molecular dynamics (MD) simulations \cite{DangicPhysRevB.106.134113}\st{,}\textcolor{red}{.}
%\textcolor{red}{Conversely,} \st{C}\textcolor{red}{c}alorimetry detects a latent heat at the transition \cite{sist2018}, implying that it is weakly first order, in agreement with \st{some} \textcolor{red}{predictions from renormalization-group-theory} calculations \cite{rabe1987}. \textcolor{red}{Finally,} EXAFS data \cite{fons2010} show an apparent survival of long and short \ce{Ge}-\ce{Te} bonds at $T\ge 700$ K, suggesting an order-disorder phase transition \cite{fons2010}. However, \st{this} \textcolor{red}{the latter} result has been reinterpreted as due to anharmonicity in the potential in Ref. \cite{DangicPhysRevB.106.134113}. \textcolor{red}{These observations seem to indicate that, in the thermal case, a first order structural phase transition slightly precedes the \it{latent} displacive phase transition.}

Femtosecond visible laser pulses induce a  non thermal transformation on GeTe from the initial rhombohedral phase to the cubic structure \cite{jianbo2015, matsubara2016}. Even if apparently  similar to the thermal one, the nature of this transition is rather controversial. A key feature is that even after 6 ps, the ultrafast X-ray diffraction (XRD) data are not consistent with a complete transition to a cubic structure as additional diffraction peaks and anomalous broadenings are still present. This led the authors of Ref. \cite{matsubara2016} to speculate that the structure after photoexcitation is only on average cubic, with the Ge atoms performing an off-center rattling motion leading to the averaging effect. This interpretation, consistent with an order to disorder phase transition, has been questioned by time-dependent  density functional theory MD calculations resulting in a coherent displacive second order phase transition  to the cubic phase  \cite{chen2018}. %However, no experimental evidence of phonon softening has ever been reported. 
%Moreover, 
However, the time evolution of the XRD lines and the persistence of some rhombohedral features are not explained by a displacive coherent  transition as proposed in Ref. \cite{chen2018}.

In this work we investigate the photoexcited phase transition in GeTe by using constrained density functional  (cDFT) and density functional perturbation theory (cDFPT) calculations  \cite{hamnn2013, perdew1997,FahyPhysRevB.65.054302,marini2021_1} coupled with the stochastic self-consistent harmonic approximation (SSCHA) \cite{monacelli2021} to treat non-perturbative quantum anharmonicity in the ground and in the excited state. The crystal under photoexcitation is here characterized by the presence of two independent thermal populations of photoexcited carriers: photoexcited electrons in the bottom of the conduction band ($\text{ph. e}^{-}$ in the following) and valence holes in the top of the valence band \cite{marini2021_1} (more details in the Supplemental Material (SM)). We show that the photoinduced non-thermal phase transition is of the first order kind and does not involve a soft phonon mode, at odd with the thermal case. The behaviour of the ultrafast XRD peaks are naturally explained in this framework.

\begin{figure}[b!]
\centering
\includegraphics[width=1\linewidth, trim={2cm 1cm 4cm 0cm},clip]{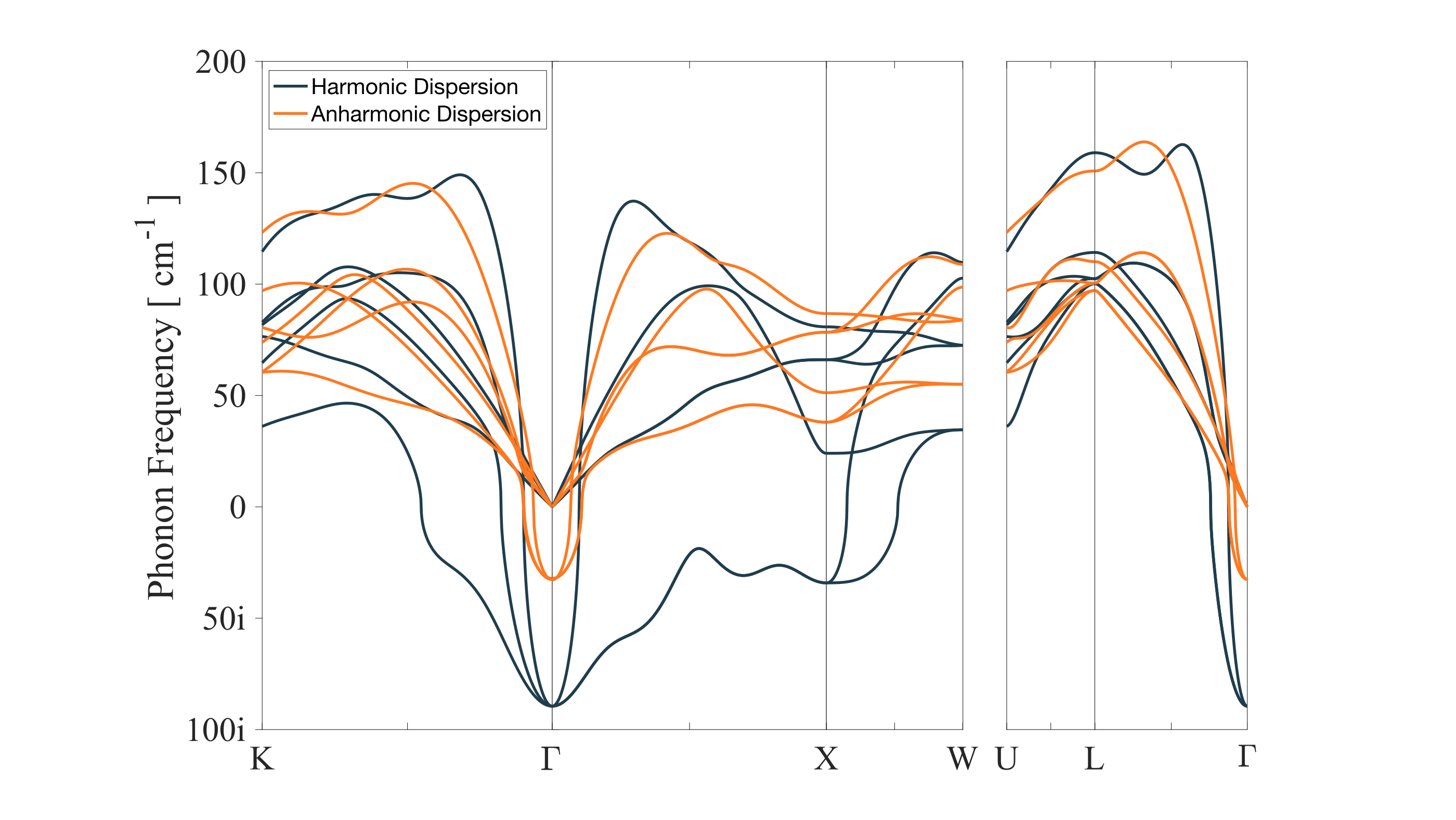}
\caption{Harmonic phonon dispersion and anharmonic $\omega_{{\bf q}\nu}$ (at $T = \SI{300}{\kelvin}$) for ground state cubic \ce{GeTe} . The high symmetry points are labeled according to the Brillouin zone of the $\text{Fm}\bar{3}m$ space group. The LO/TO splitting is not included in the calculation.}
\label{fig3}
\end{figure}

We first consider the thermal case and show that in the absence of photoexcitation the SSCHA correctly reproduces the displacive nature of the instability of the cubic phase at $T=300$ K.
In Fig. \ref{fig3}, we show the harmonic phonon dispersion of $\beta$-\ce{GeTe} calculated by using density functional perturbation theory \cite{baroni2001,QE,QE2}. In the harmonic approximation, the phase is dynamically unstable, with the complete optical band being imaginary and with the strongest instability located at zone center. We explicitly verify that the inclusion of the natural hole concentration due to the intrinsic Ge vacancies in \ce{GeTe} increases the harmonic optical phonon frequency by no more than $5\%$; thus, in the following, their presence will be neglected (see SM and Ref. \cite{edwards2005} there included).\\
Non-perturbative quantum anharmonicity is treated within the SSCHA by calculating the positional free energy ($F$) Hessian at $T=300$ K. The static temperature dependent dynamical matrix is defined as a function of the free energy Hessian as:
\begin{equation}
    \textbf{\textit{D}} = \textbf{M}^{-\frac{1}{2}}\frac{\partial^2 F}{\partial \textbf{R}^2} \Bigg| _{\textbf{R}_{eq}} \textbf{M}^{-\frac{1}{2}}
\end{equation}
where $\textbf{M}$ is the matrix of the ionic masses $M_a$ with $M_{ab}$ = $\delta_{ab} M_a$ and $\textbf{R}$ is a cumulative variable for all the ionic positions (see Refs. \onlinecite{bianco2017second} for a detailed explanation). By  Fourier transforming the matrix $\textbf{\textit{D}}$ and  by diagonalizing it, we obtain $\omega_{{\bf q}\nu}$ as the  square root of the eigenvalues of the positional free energy Hessian. An imaginary frequency signals the occurrence of a second order phase transition as dictated by Landau theory.  
As shown in Fig. \ref{fig3}, quantum anharmonicity
partially cures the instability as most of the optical band is now positive, but there remains an instability of the optical mode at zone center indicating the occurrence of a displacive phase transition, in  agreement with experiments and  previous calculations. \cite{DangicPhysRevB.106.134113}.

\begin{figure}[t!]
\centering
\includegraphics[width=1.05\linewidth, trim={1.2cm 0.5cm 1cm .8cm},clip]{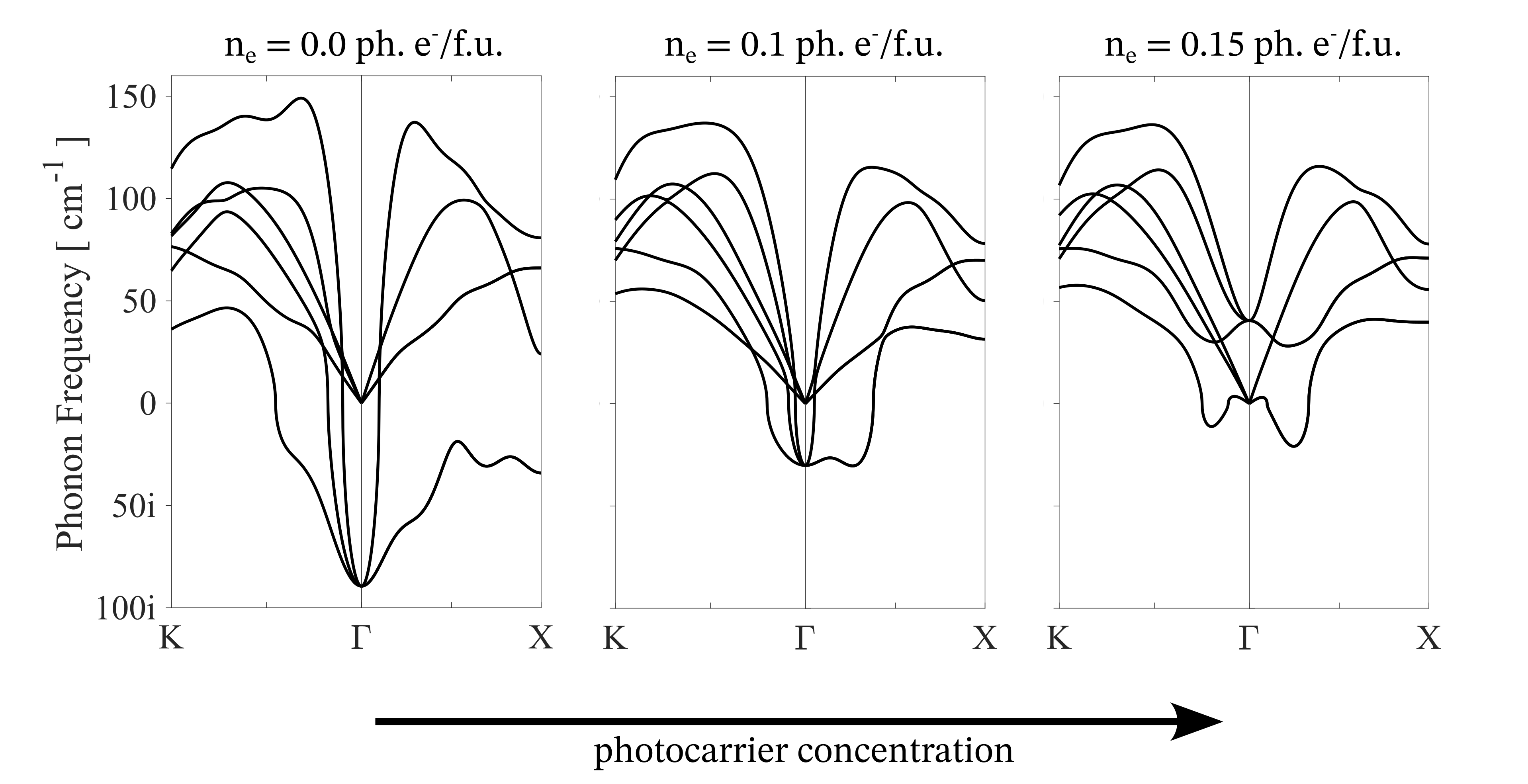}
\caption{Harmonic phonon dispersion for cubic \ce{GeTe} as a function of increasing photocarrier concentration. The LO/TO splitting is not included in the calculation.}
\label{fig2}
\end{figure}

Having shown the capability of our formalism to describe the displacive phase transition in the ground state, we switch to the description of the photoexcited state. In  Fig. \ref{fig5} (c) we show the harmonic phonon dispersion of the $\alpha$-phase in the ground state and in the presence of $n_e = 0.1$ excited electrons per formula unit ($\text{ph.}\, \text{e}^- / \text{ f.u.}$), which is compatible with the value that we calculate from the experimental fluence \cite{matsubara2016} (see SM which includes Ref. \cite{lewis1973}). The ground state results are in perfect agreement with Refs. \cite{wdowik2014,shltaf2008,CampiPhysRevB.95.024311} and confirm the stability of the $\alpha$-phase at zero temperature. More interestingly, the rhombohedral phase remains stable even in the presence of a substantial photoexcitation and shows no tendency to a dynamical destabilization of the lattice. Quantum anharmonicity weakly affects the dispersion at $n_e=0.1 \, \text{ph.}\, \text{e}^- / \text{ f.u.}$, as show in Fig. \ref{fig5} (d). 

\begin{figure}[t!]
\centering
\includegraphics[width=1\linewidth, trim={2cm 1cm 4cm 0cm},clip]{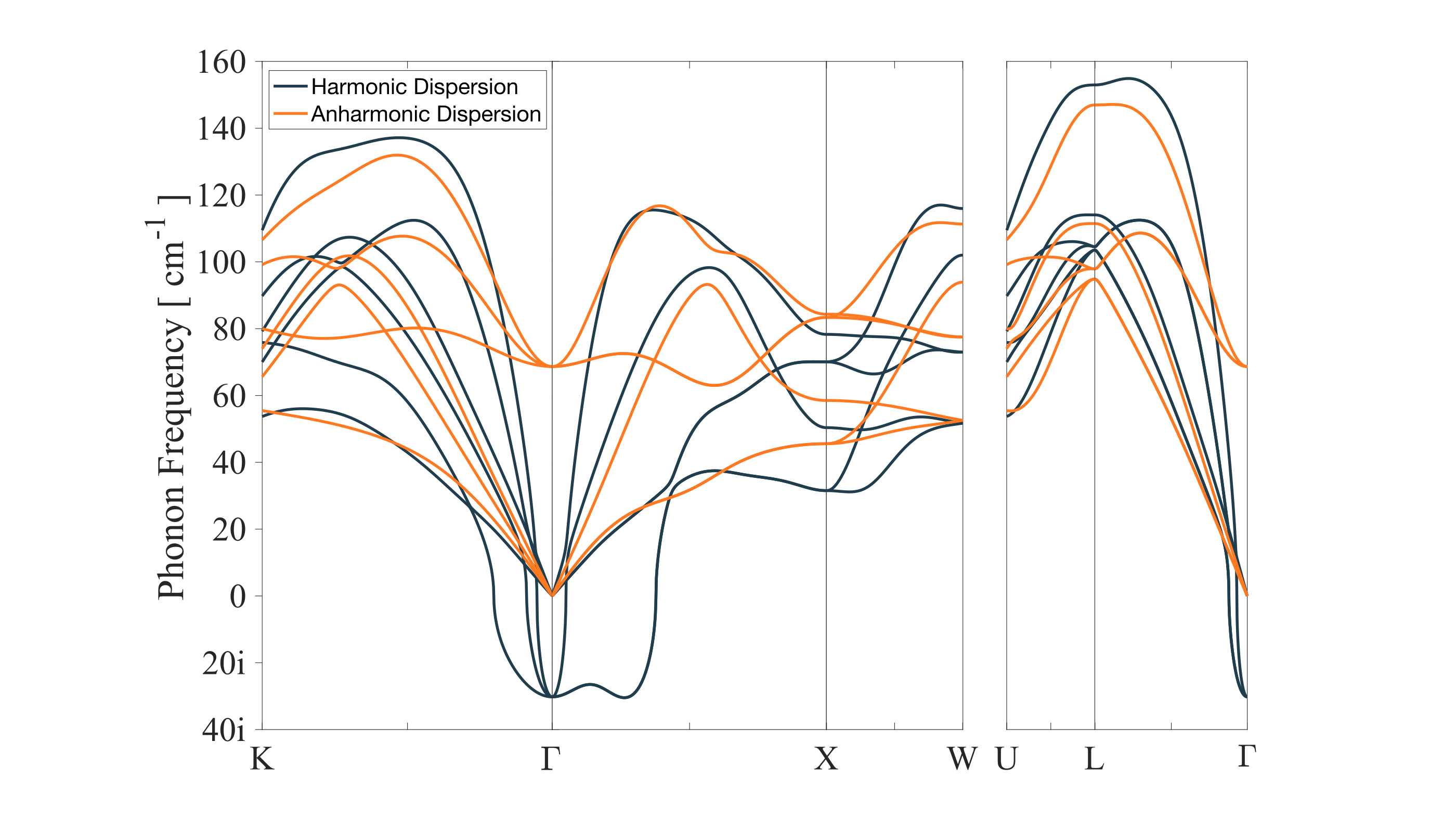}
\caption{Harmonic and anharmonic phonon dispersion (at $T = \SI{300}{\kelvin}$)  for photoexcited cubic \ce{GeTe} at $n_\text{e} = 0.1 \,  \text{ph.}\, \text{e}^-/\text{f.u.}$. The high symmetry points are labeled according to the Brillouin zone of the $\text{Fm}\bar{3}m$ space group.}
\label{fig4}
\end{figure}

\begin{figure*}[t]
\centering
\includegraphics[width=1\linewidth, trim={0cm .0cm 0cm 0cm},clip]{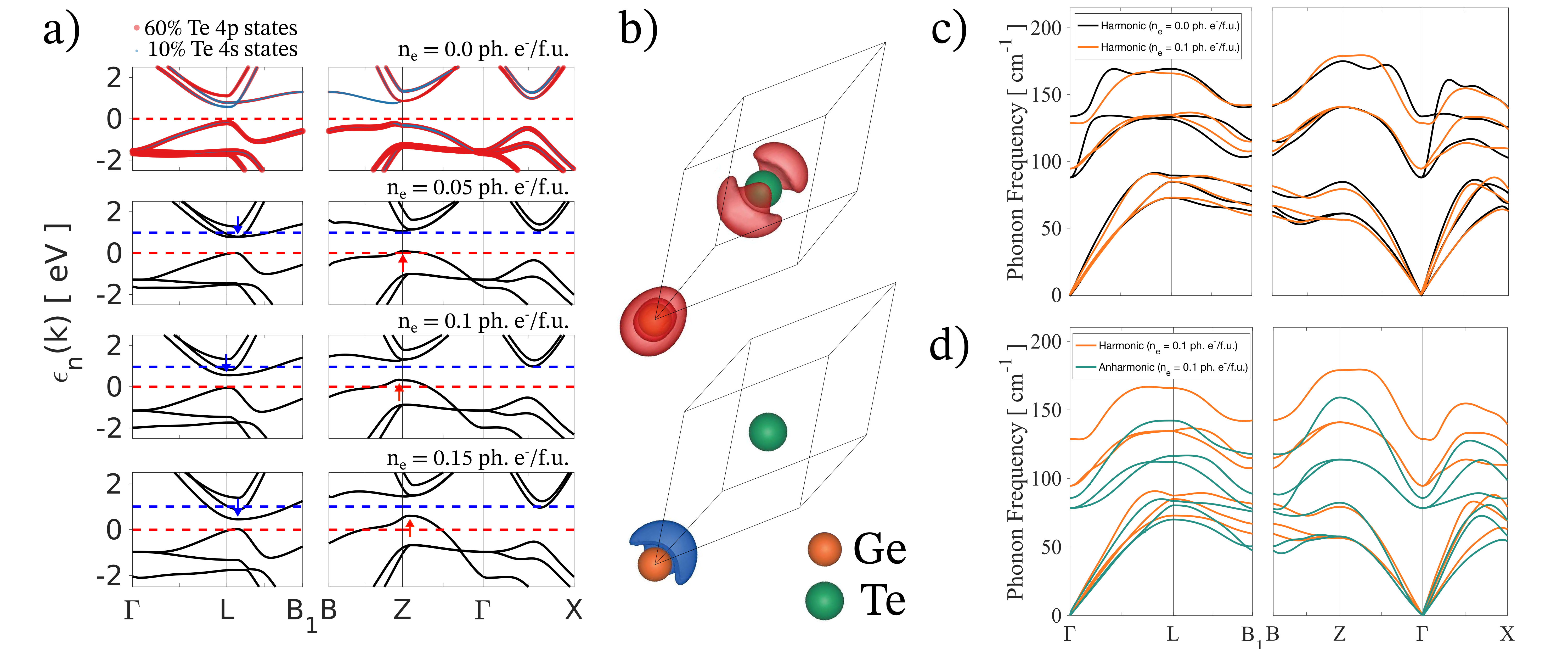}
\caption{a) From top to bottom, electronic band structure of $\alpha$-\ce{GeTe} for increasing concentrations of photoexcited electrons; the atomic state projections on \ce{Te} \textit{s} and \textit{p} states are superimposed on the ground state electronic band structure; the Kohn-Sham eigenvalues are referred to the valence hole Fermi level (dashed blue line), the photoexcited electron Fermi level is reported with a dashed red line instead. b) Spatial distribution of electronic charge density difference in $\alpha$-\ce{GeTe} assuming a p-doping of  $0.1 \text{ h}^- / \text{ f.u.}$, (top) and a n-doping of $0.1 \text{ e}^- / \text{ f.u.}$ (bottom). In both cases only the $3.5 \, \text{e}^- /\text{\small{\AA}}^3$ isosurface is shown \cite{Momma:db5098}. Harmonic (c and d) phonon dispersion and anharmonic $\omega_{{\bf q} \nu}$ (at $T = \SI{300}{\kelvin}$) (d)  of (c) ground state and (d) photoexcited ($0.1 \, \, \text{ph. e}^- / \text{ f.u.}$) $\alpha$-\ce{GeTe}. The high symmetry points are labelled according to the Brillouin zone of the R3m space group following Ref. \cite{Setyawan2010} and the k-space paths were geenrated through the Xcrysden software \cite{KOKALJ1999176}.}
\label{fig5}
\end{figure*}

It is instructive to compare the persistent stability of the $\alpha$-\ce{GeTe} phase under photoexcitation with the  harmonic phonon dispersion of the cubic phase as a function of the number of photoexcited carriers (Fig. \ref{fig2}). A non-zero photoexcited carriers population progressively reduces the displacive instability of the optical phonons at zone center. At the harmonic level and in the absence of photocarriers cubic \ce{GeTe} exhibits a deep dynamical instability at zone centre ($\omega = 89\,i \text{ } \SI{}{\per \cm}$) and a secondary dynamical instability at $X$ ($\omega = 34\,i \text{ } \SI{}{\per \cm}$), in complete agreement with previous theoretical determination of the same vibrational spectrum \cite{wdowik2014, xia2018, wang2021}.
At  $n_e=0.1 \,  \text{ph.}\, \text{e}^- / \text{ f.u.}$  the instability of the zone-center optical-phonon is weak but still present, while it is completely removed at $n_e=0.15 \,  \text{ph.}\, \text{e}^- / \text{ f.u.}$, although the acoustic modes are slightly imaginary close to zone center. We note here that photoinduced weakening of the zone centre dynamical instability can be correlated to the promotion of electrons to the conduction bands of $\beta$-\ce{GeTe} which are predominantly derived from Ge p states and thus favour a more perfect octahedral atomic arrangement (see SM for the $\beta$-GeTe band character) \cite{wuttig2007}.\\ 
The residual weak instability at $n_e \ge 0.1 \, \text{ph.}\, \text{e}^- / \text{ f.u.}$ could, however, be removed by anharmonicity. In order to investigate this possibility we report the results of the  SSCHA calculation at $T=\SI{300}{\kelvin}$ and $n_e=0.1 \, \text{ph.}\, \text{e}^- / \text{ f.u.}$ in Fig. \ref{fig4}. Quantum anharmonic corrections are colossal and completely remove the instability. Thus, both the $\alpha$ and $\beta$ phases have a positive definite positional free energy Hessian at this photoexcitation, meaning that both phases are local free energy minima and the transition among the two cannot be due to a displacive soft-phonon driven phase transition, at odd with the thermal case. Inclusion of volume effects and minimization of the Gibbs free energy do not qualitatively and quantitatively change this conclusion. %Finally, in the SM we also report the spectral weight showing the occurrence of a phonon-satellite at zone center. 
%uncertainty1 0.1 and uncertainty2 0.07
At $n_\text{e} = 0.1 \,  \text{ph.}\, \text{e}^-/\text{f.u.}$ the quantum free energy difference between the cubic and the rhombohedral phases is very small, namely $\Delta F \vcentcolon=  F_{\beta-\ce{GeTe}} - F_{\alpha-\ce{GeTe}} = \SI{2.4}{\milli \eV}/\text{f.u.}$ at T = \SI{300}{\kelvin}. Inclusion of volume effects and minimization of the Gibbs free energy changes only slightly this result and leads to a quantum free energy difference of $\Delta F = \SI{2.36}{\milli \eV}/\text{f.u.}$ at $T=\SI{300}{\kelvin}$ (the total energy difference being $\Delta E = \SI{16}{\milli \eV}/\text{f.u.}$ ). This has to be compared with the quantum free energy difference in the absence of photoexcitation at $T=\SI{300}{\kelvin}$, that is $\Delta F = \SI{18}{\milli \eV}/\text{f.u.}$ (the total energy difference in the ground state is $\Delta E = \SI{29}{\milli \eV}$). Thus, photoexcitation suppresses the free energy difference among the two structures by a factor $7.6$, making them almost degenerate at $n_e=0.1 \,  \text{ph.}\, \text{e}^-/ \text{ f.u.}$.
The reduction in $\Delta F$ is entirely due to the change in total energy difference by the photoexcitation, while  the quantum anharmonic correction is almost equal in the ground state and in the excited state. Our results obtained within cDFPT strongly support the occurrence of a first order phase transition from the rhombohedral to the cubic phase with a region of coexistence among the two.

Before addressing the comparison with XRD data, we detail the microscopic mechanism responsible for the free energy difference reduction under photoexcitation. In Fig. \ref{fig5} (a) we plot the evolution of the  electronic structure of the $\alpha$-phase as a function of photoexcitation obtained with cDFT. As shown by the projection over the Te atomic character, in the ground state the valence band has a dominant Te character (at the valence band top approximately $60\%$ of Te and $40\%$ of Ge character), while the conduction band is mostly derived from Ge atomic states. This is also shown clearly in Fig. \ref{fig5} (b), where the electronic charge density restricted to the top of the valence band (top) and to the bottom of the conduction band (bottom)  is shown (see SM for more details).  The former is almost equally distributed among the Ge and Te atoms, while the latter is mostly located close to the Ge atoms.
As Ge (Te) is positively (negatively) charged in the ground state \cite{Singh_Sci_Rep,shltaf2008}, the photoexcitation reduces the ionicity of the bonding. As the stabilization of the rhombohedral structure in the ground state has been attributed to the smaller ionicity of the bonding \cite{wuttig2007, shportko2008,shltaf2008} and the consequent  energy gain due to gap-opening,  the  stabilization of the cubic structure in the presence of photocarriers seems to be in apparent contradiction with previous works. However, we should recall that at $n_e=0.1 \,  \text{ph.}\, \text{e}^-/{f.u.}$, GeTe is in a metallic state. The concept of bond ionicity is lost as the Born effective charges are completely screened, no net polarization occurs in the $\alpha-$phase and the band-gap is progressively reduced, as it can be seen in Fig. \ref{fig5}. 
On the contrary the gap of the cubic phase is essentially unaltered by photoexcitation (see SM).
Thus, the reduction in $\Delta F$ and $\Delta E$ by the electron-hole plasma is mostly due to the gap closure.

%Finally, it is interesting to remark that within cDFT the increased photoexcited carrier concentration induces a direct to indirect gap transition and finally a transition to a metallic state at $n_e\approx 0.15$. This is a common feature of photoexcited insulators \cite{marini2021_2}.
%

\begin{figure*}[t!]
\centering
\includegraphics[width=\linewidth]{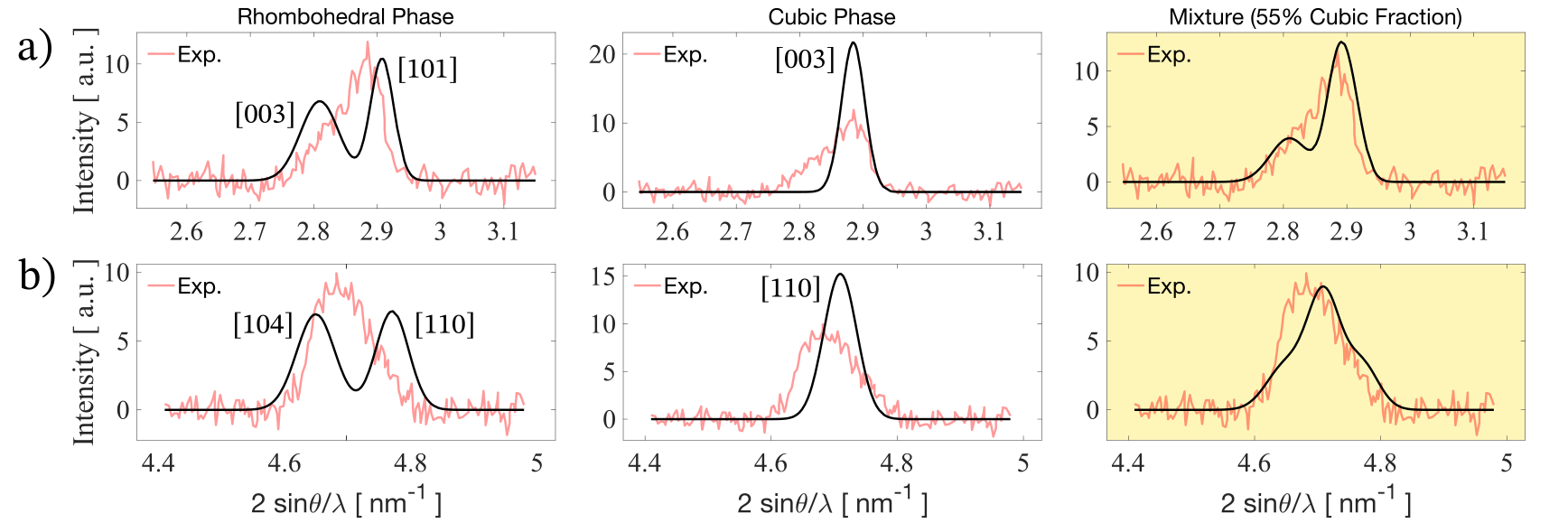}
\caption{Comparison between experimental time resolved XRD data  ($\SI{6}{\pico \second}$ after photoexcitation,  replotted as Exp. from Ref. \cite{matsubara2016}) and the calculated diffraction peaks of $\alpha$-\ce{GeTe}, $\beta$-\ce{GeTe} and a mixture of the two phases ($55\%$ $\beta$-\ce{GeTe} concentration), respectively. In the top row the comparison focuses on the range where the $[003]$ and $[101]$ rhombohedral peaks are visible in Ref. \cite{matsubara2016}. In the bottom row, the comparison is limited over the range where the $[104]$ and $[110]$ rhombohedral peaks are found in Ref. \cite{matsubara2016} instead. All intensities are normalized to unity.}
\label{fig6}
\end{figure*}

We now compare our findings with experimental data. Ultrafast XRD data in Ref. \cite{matsubara2016} show a progressive emergence of the cubic peaks in the time period 4-26 \SI{}{\pico \second} after the transition. Yet, the behaviour of the  $[003]$, $[104]$ and $[110]$ peaks is anomalous, as shown in  Fig. \ref{fig6}. In particular, in the cubic phase, only the $[003]$ peak should be visible in the range of $2.6< 2 \sin\theta/\lambda<3$ nm$^{-1}$, as shown in the simulated diffraction pattern (black line) in the middle panel in Fig. \ref{fig6} (a). However, at 6 ps the experimental data of Ref. \cite{matsubara2016}, replotted in Fig. \ref{fig6} (a), show the presence of a secondary  peak as a left-shoulder.  Similarly, in the case of a complete transition to $\beta$-\ce{GeTe}, the cubic $[110]$ peak should be the only visible in the diffraction pattern in the range $4.5< 2 \sin\theta/\lambda<4.9$ nm$^{-1}$. On the contrary, the asymmetric shape of the experimental peak suggests a persistence of the rhombohedral $[104]$ at 6 ps (and also at larger times), see middle panel of Fig. \ref{fig6} (b).

These anomalies in the XRD pattern were interpreted as a rattling motion of the central ion in the unit cell that  experiences a shallow enough potential energy barrier at the geometric centre of the unit cell so to be able to rapidly move around it and leading, on average, to a cubic structure \cite{matsubara2016}.
Although such an explanation was suggested to be consistent with the experimental observations through theoretical estimation of the structure factors in rhombohedral \ce{GeTe}, no clear evidence of this specific dynamics was provided. Such a scenario clearly differs significantly from the first order picture of the phase transition provided here. We believe that the progressive change in the diffraction patterns in Ref. \cite{matsubara2016} could be more conveniently explained assuming that after the irradiation a certain portion of the sample transitions to the $\beta$ phase. The larger concentration of $\beta$-\ce{GeTe} in the sample would then correlate with the observed enhancement of the diffraction peaks pertaining to the cubic structure. The remaining portion still in the $\alpha$ phase would naturally give rise to additional features in the XRD patterns.

In Fig. \ref{fig6}, we provide a comparison of the diffraction peaks reported in Ref. \cite{matsubara2016} with theoretically computed diffraction patterns \cite{Momma:db5098}. For simplicity, we chose the linewidths of the theoretical peaks to be identical to the experimental ones for ground state $\alpha$-\ce{GeTe} (see Ref. \cite{matsubara2016} and SM for further details). In Fig. \ref{fig6} it is evident how purely rhombohedral or purely cubic diffraction patterns cannot explain the experimental observations. However, when considering mixtures of the two phases, the qualitative features of the experimental diffraction patterns are recovered (see \ref{fig6} and SM for more details). Thus, the XRD in Ref. \cite{matsubara2016} are better described by a phase mixture than by a second-order displacive transition. 
 
In conclusion, by using first principles calculations in the presence of an electron-hole plasma and by including non-perturbative  quantum anharmonicity, we showed that the non-thermal rhombohedral to cubic phase transition occurring in photoexcited GeTe is of the first order kind and does not involve softening of any optical zone centre mode, contrary to what happens in the thermal case. By performing a careful analysis of ultrafast XRD data \cite{matsubara2016}, we have shown that they are compatible with a mixture of the $\alpha$ and $\beta$ phases and disagree with a second-order or with a coherently uniform transition to the $\beta-$phase \cite{chen2018}.\\
Our results are relevant for the broad class of phase change materials, as in several of these, such as SbTe$_3$ or Ge$_2$Sb$_2$Te$_5$, non-thermal phase transition induced by ultrafast pulses have been detected \cite{Hase2015}. Moreover, as we have shown that the order of the transition is sensitive to the effective occupation of the valence band, our findings could be helpful to explain the weakly first order nature of the thermal transition in \ce{GeTe} and other phase change materials. 

\acknowledgments
We acknowledge EuroHPC access on LUMI (EHPC-REG-2022R03-090) for high performance computing resources and the CINECA award under the ISCRA initiative, for the availability of high-performance computing resources and support. This work was funded by the European Union (ERC, DELIGHT, 101052708). Views and opinions expressed are however those of the author(s) only and do not necessarily reflect those of the European Union or the European Research Council. Neither the European Union nor the granting authority can be held responsible for them.


\begin{thebibliography}{38}%
\makeatletter
\providecommand \@ifxundefined [1]{%
 \@ifx{#1\undefined}
}%
\providecommand \@ifnum [1]{%
 \ifnum #1\expandafter \@firstoftwo
 \else \expandafter \@secondoftwo
 \fi
}%
\providecommand \@ifx [1]{%
 \ifx #1\expandafter \@firstoftwo
 \else \expandafter \@secondoftwo
 \fi
}%
\providecommand \natexlab [1]{#1}%
\providecommand \enquote  [1]{``#1''}%
\providecommand \bibnamefont  [1]{#1}%
\providecommand \bibfnamefont [1]{#1}%
\providecommand \citenamefont [1]{#1}%
\providecommand \href@noop [0]{\@secondoftwo}%
\providecommand \href [0]{\begingroup \@sanitize@url \@href}%
\providecommand \@href[1]{\@@startlink{#1}\@@href}%
\providecommand \@@href[1]{\endgroup#1\@@endlink}%
\providecommand \@sanitize@url [0]{\catcode `\\12\catcode `\$12\catcode
  `\&12\catcode `\#12\catcode `\^12\catcode `\_12\catcode `\%12\relax}%
\providecommand \@@startlink[1]{}%
\providecommand \@@endlink[0]{}%
\providecommand \url  [0]{\begingroup\@sanitize@url \@url }%
\providecommand \@url [1]{\endgroup\@href {#1}{\urlprefix }}%
\providecommand \urlprefix  [0]{URL }%
\providecommand \Eprint [0]{\href }%
\providecommand \doibase [0]{https://doi.org/}%
\providecommand \selectlanguage [0]{\@gobble}%
\providecommand \bibinfo  [0]{\@secondoftwo}%
\providecommand \bibfield  [0]{\@secondoftwo}%
\providecommand \translation [1]{[#1]}%
\providecommand \BibitemOpen [0]{}%
\providecommand \bibitemStop [0]{}%
\providecommand \bibitemNoStop [0]{.\EOS\space}%
\providecommand \EOS [0]{\spacefactor3000\relax}%
\providecommand \BibitemShut  [1]{\csname bibitem#1\endcsname}%
\let\auto@bib@innerbib\@empty
%</preamble>
\bibitem [{\citenamefont {Zhang}\ \emph {et~al.}(2019)\citenamefont {Zhang},
  \citenamefont {Mazzarello}, \citenamefont {Wuttig},\ and\ \citenamefont
  {Ma}}]{Zhang2019}%
  \BibitemOpen
  \bibfield  {author} {\bibinfo {author} {\bibfnamefont {W.}~\bibnamefont
  {Zhang}}, \bibinfo {author} {\bibfnamefont {R.}~\bibnamefont {Mazzarello}},
  \bibinfo {author} {\bibfnamefont {M.}~\bibnamefont {Wuttig}},\ and\ \bibinfo
  {author} {\bibfnamefont {E.}~\bibnamefont {Ma}},\ }\href@noop {} {\bibfield
  {journal} {\bibinfo  {journal} {Nature Reviews Materials}\ }\textbf {\bibinfo
  {volume} {4}},\ \bibinfo {pages} {150} (\bibinfo {year} {2019})}\BibitemShut
  {NoStop}%
\bibitem [{\citenamefont {Ovshinsky}(1968)}]{ovshinsky1968}%
  \BibitemOpen
  \bibfield  {author} {\bibinfo {author} {\bibfnamefont {S.~R.}\ \bibnamefont
  {Ovshinsky}},\ }\href@noop {} {\bibfield  {journal} {\bibinfo  {journal}
  {Phys. Rev. Lett.}\ }\textbf {\bibinfo {volume} {21}},\ \bibinfo {pages}
  {1450} (\bibinfo {year} {1968})}\BibitemShut {NoStop}%
\bibitem [{\citenamefont {Lencer}\ \emph {et~al.}(2008)\citenamefont {Lencer},
  \citenamefont {Salinga}, \citenamefont {Grabowski}, \citenamefont {Hickel},
  \citenamefont {Neugebauer},\ and\ \citenamefont {Wuttig}}]{lencer2008}%
  \BibitemOpen
  \bibfield  {author} {\bibinfo {author} {\bibfnamefont {D.}~\bibnamefont
  {Lencer}}, \bibinfo {author} {\bibfnamefont {M.}~\bibnamefont {Salinga}},
  \bibinfo {author} {\bibfnamefont {B.}~\bibnamefont {Grabowski}}, \bibinfo
  {author} {\bibfnamefont {T.}~\bibnamefont {Hickel}}, \bibinfo {author}
  {\bibfnamefont {J.}~\bibnamefont {Neugebauer}},\ and\ \bibinfo {author}
  {\bibfnamefont {M.}~\bibnamefont {Wuttig}},\ }\href@noop {} {\bibfield
  {journal} {\bibinfo  {journal} {Nature Materials}\ }\textbf {\bibinfo
  {volume} {7}},\ \bibinfo {pages} {972} (\bibinfo {year} {2008})}\BibitemShut
  {NoStop}%
\bibitem [{\citenamefont {Wuttig}\ \emph {et~al.}(2017)\citenamefont {Wuttig},
  \citenamefont {Bhaskaran},\ and\ \citenamefont {Taubner}}]{Wuttig2017}%
  \BibitemOpen
  \bibfield  {author} {\bibinfo {author} {\bibfnamefont {M.}~\bibnamefont
  {Wuttig}}, \bibinfo {author} {\bibfnamefont {H.}~\bibnamefont {Bhaskaran}},\
  and\ \bibinfo {author} {\bibfnamefont {T.}~\bibnamefont {Taubner}},\ }\href
  {https://doi.org/10.1038/nphoton.2017.126} {\bibfield  {journal} {\bibinfo
  {journal} {Nature Photonics}\ }\textbf {\bibinfo {volume} {11}},\ \bibinfo
  {pages} {465} (\bibinfo {year} {2017})}\BibitemShut {NoStop}%
\bibitem [{\citenamefont {Shportko}\ \emph {et~al.}(2008)\citenamefont
  {Shportko}, \citenamefont {Kremers}, \citenamefont {Woda}, \citenamefont
  {Lencer}, \citenamefont {Robertson},\ and\ \citenamefont
  {Wuttig}}]{shportko2008}%
  \BibitemOpen
  \bibfield  {author} {\bibinfo {author} {\bibfnamefont {K.}~\bibnamefont
  {Shportko}}, \bibinfo {author} {\bibfnamefont {S.}~\bibnamefont {Kremers}},
  \bibinfo {author} {\bibfnamefont {M.}~\bibnamefont {Woda}}, \bibinfo {author}
  {\bibfnamefont {D.}~\bibnamefont {Lencer}}, \bibinfo {author} {\bibfnamefont
  {J.}~\bibnamefont {Robertson}},\ and\ \bibinfo {author} {\bibfnamefont
  {M.}~\bibnamefont {Wuttig}},\ }\href@noop {} {\bibfield  {journal} {\bibinfo
  {journal} {Nature Materials}\ }\textbf {\bibinfo {volume} {7}},\ \bibinfo
  {pages} {653} (\bibinfo {year} {2008})}\BibitemShut {NoStop}%
\bibitem [{\citenamefont {Matsubara}\ \emph {et~al.}(2016)\citenamefont
  {Matsubara}, \citenamefont {Okada}, \citenamefont {Ichitsubo}, \citenamefont
  {Kawaguchi}, \citenamefont {Hirata}, \citenamefont {Guan}, \citenamefont
  {Tokuda}, \citenamefont {Tanimura}, \citenamefont {Matsunaga}, \citenamefont
  {Chen},\ and\ \citenamefont {Yamada}}]{matsubara2016}%
  \BibitemOpen
  \bibfield  {author} {\bibinfo {author} {\bibfnamefont {E.}~\bibnamefont
  {Matsubara}}, \bibinfo {author} {\bibfnamefont {S.}~\bibnamefont {Okada}},
  \bibinfo {author} {\bibfnamefont {T.}~\bibnamefont {Ichitsubo}}, \bibinfo
  {author} {\bibfnamefont {T.}~\bibnamefont {Kawaguchi}}, \bibinfo {author}
  {\bibfnamefont {A.}~\bibnamefont {Hirata}}, \bibinfo {author} {\bibfnamefont
  {P.~F.}\ \bibnamefont {Guan}}, \bibinfo {author} {\bibfnamefont
  {K.}~\bibnamefont {Tokuda}}, \bibinfo {author} {\bibfnamefont
  {K.}~\bibnamefont {Tanimura}}, \bibinfo {author} {\bibfnamefont
  {T.}~\bibnamefont {Matsunaga}}, \bibinfo {author} {\bibfnamefont {M.~W.}\
  \bibnamefont {Chen}},\ and\ \bibinfo {author} {\bibfnamefont
  {N.}~\bibnamefont {Yamada}},\ }\href@noop {} {\bibfield  {journal} {\bibinfo
  {journal} {Phys. Rev. Lett.}\ }\textbf {\bibinfo {volume} {117}},\ \bibinfo
  {pages} {135501} (\bibinfo {year} {2016})}\BibitemShut {NoStop}%
\bibitem [{\citenamefont {Boschker}\ \emph {et~al.}(2017)\citenamefont
  {Boschker}, \citenamefont {Wang},\ and\ \citenamefont
  {Calarco}}]{boschker2017}%
  \BibitemOpen
  \bibfield  {author} {\bibinfo {author} {\bibfnamefont {J.~E.}\ \bibnamefont
  {Boschker}}, \bibinfo {author} {\bibfnamefont {R.}~\bibnamefont {Wang}},\
  and\ \bibinfo {author} {\bibfnamefont {R.}~\bibnamefont {Calarco}},\
  }\href@noop {} {\bibfield  {journal} {\bibinfo  {journal} {CrystEngComm}\
  }\textbf {\bibinfo {volume} {19}},\ \bibinfo {pages} {5324} (\bibinfo {year}
  {2017})}\BibitemShut {NoStop}%
\bibitem [{\citenamefont {Chattopadhyay}\ \emph {et~al.}(1987)\citenamefont
  {Chattopadhyay}, \citenamefont {Boucherle},\ and\ \citenamefont
  {vonSchnering}}]{chattopadhyay1987}%
  \BibitemOpen
  \bibfield  {author} {\bibinfo {author} {\bibfnamefont {T.}~\bibnamefont
  {Chattopadhyay}}, \bibinfo {author} {\bibfnamefont {J.~X.}\ \bibnamefont
  {Boucherle}},\ and\ \bibinfo {author} {\bibfnamefont {H.~G.}\ \bibnamefont
  {vonSchnering}},\ }\href@noop {} {\bibfield  {journal} {\bibinfo  {journal}
  {Journal of Physics C: Solid State Physics}\ }\textbf {\bibinfo {volume}
  {20}},\ \bibinfo {pages} {1431} (\bibinfo {year} {1987})}\BibitemShut
  {NoStop}%
\bibitem [{\citenamefont {Chatterji}\ \emph {et~al.}(2015)\citenamefont
  {Chatterji}, \citenamefont {Kumar},\ and\ \citenamefont
  {Wdowik}}]{chatterji2015}%
  \BibitemOpen
  \bibfield  {author} {\bibinfo {author} {\bibfnamefont {T.}~\bibnamefont
  {Chatterji}}, \bibinfo {author} {\bibfnamefont {C.~M.~N.}\ \bibnamefont
  {Kumar}},\ and\ \bibinfo {author} {\bibfnamefont {U.~D.}\ \bibnamefont
  {Wdowik}},\ }\href@noop {} {\bibfield  {journal} {\bibinfo  {journal} {Phys.
  Rev. B}\ }\textbf {\bibinfo {volume} {91}},\ \bibinfo {pages} {054110}
  (\bibinfo {year} {2015})}\BibitemShut {NoStop}%
\bibitem [{\citenamefont {Steigmeier}\ and\ \citenamefont
  {Harbeke}(1970)}]{stegmeier1970}%
  \BibitemOpen
  \bibfield  {author} {\bibinfo {author} {\bibfnamefont {E.~F.}\ \bibnamefont
  {Steigmeier}}\ and\ \bibinfo {author} {\bibfnamefont {G.}~\bibnamefont
  {Harbeke}},\ }\href@noop {} {\bibfield  {journal} {\bibinfo  {journal} {Solid
  State Communications}\ }\textbf {\bibinfo {volume} {8}},\ \bibinfo {pages}
  {1275} (\bibinfo {year} {1970})}\BibitemShut {NoStop}%
\bibitem [{\citenamefont {Sosso}\ \emph {et~al.}(2012)\citenamefont {Sosso},
  \citenamefont {Miceli}, \citenamefont {Caravati}, \citenamefont {Behler},\
  and\ \citenamefont {Bernasconi}}]{Sosso2012}%
  \BibitemOpen
  \bibfield  {author} {\bibinfo {author} {\bibfnamefont {G.~C.}\ \bibnamefont
  {Sosso}}, \bibinfo {author} {\bibfnamefont {G.}~\bibnamefont {Miceli}},
  \bibinfo {author} {\bibfnamefont {S.}~\bibnamefont {Caravati}}, \bibinfo
  {author} {\bibfnamefont {J.}~\bibnamefont {Behler}},\ and\ \bibinfo {author}
  {\bibfnamefont {M.}~\bibnamefont {Bernasconi}},\ }\href@noop {} {\bibfield
  {journal} {\bibinfo  {journal} {Phys. Rev. B}\ }\textbf {\bibinfo {volume}
  {85}},\ \bibinfo {pages} {174103} (\bibinfo {year} {2012})}\BibitemShut
  {NoStop}%
\bibitem [{\citenamefont {Dangi\ifmmode~\acute{c}\else \'{c}\fi{}}\ \emph
  {et~al.}(2022)\citenamefont {Dangi\ifmmode~\acute{c}\else \'{c}\fi{}},
  \citenamefont {Fahy},\ and\ \citenamefont {Savi\ifmmode~\acute{c}\else
  \'{c}\fi{}}}]{DangicPhysRevB.106.134113}%
  \BibitemOpen
  \bibfield  {author} {\bibinfo {author} {\bibfnamefont {D.}~\bibnamefont
  {Dangi\ifmmode~\acute{c}\else \'{c}\fi{}}}, \bibinfo {author} {\bibfnamefont
  {S.}~\bibnamefont {Fahy}},\ and\ \bibinfo {author} {\bibfnamefont
  {I.}~\bibnamefont {Savi\ifmmode~\acute{c}\else \'{c}\fi{}}},\ }\href
  {https://doi.org/10.1103/PhysRevB.106.134113} {\bibfield  {journal} {\bibinfo
   {journal} {Phys. Rev. B}\ }\textbf {\bibinfo {volume} {106}},\ \bibinfo
  {pages} {134113} (\bibinfo {year} {2022})}\BibitemShut {NoStop}%
\bibitem [{\citenamefont {Sist}\ \emph {et~al.}(2018)\citenamefont {Sist},
  \citenamefont {Kasai}, \citenamefont {Hedegaard},\ and\ \citenamefont
  {Iversen}}]{sist2018}%
  \BibitemOpen
  \bibfield  {author} {\bibinfo {author} {\bibfnamefont {M.}~\bibnamefont
  {Sist}}, \bibinfo {author} {\bibfnamefont {H.}~\bibnamefont {Kasai}},
  \bibinfo {author} {\bibfnamefont {E.~M.~J.}\ \bibnamefont {Hedegaard}},\ and\
  \bibinfo {author} {\bibfnamefont {B.~B.}\ \bibnamefont {Iversen}},\
  }\href@noop {} {\bibfield  {journal} {\bibinfo  {journal} {Phys. Rev. B}\
  }\textbf {\bibinfo {volume} {97}},\ \bibinfo {pages} {094116} (\bibinfo
  {year} {2018})}\BibitemShut {NoStop}%
\bibitem [{\citenamefont {Fons}\ \emph {et~al.}(2010)\citenamefont {Fons},
  \citenamefont {Kolobov}, \citenamefont {Krbal}, \citenamefont {Tominaga},
  \citenamefont {Andrikopoulos}, \citenamefont {Yannopoulos}, \citenamefont
  {Voyiatzis},\ and\ \citenamefont {Uruga}}]{fons2010}%
  \BibitemOpen
  \bibfield  {author} {\bibinfo {author} {\bibfnamefont {P.}~\bibnamefont
  {Fons}}, \bibinfo {author} {\bibfnamefont {A.~V.}\ \bibnamefont {Kolobov}},
  \bibinfo {author} {\bibfnamefont {M.}~\bibnamefont {Krbal}}, \bibinfo
  {author} {\bibfnamefont {J.}~\bibnamefont {Tominaga}}, \bibinfo {author}
  {\bibfnamefont {K.~S.}\ \bibnamefont {Andrikopoulos}}, \bibinfo {author}
  {\bibfnamefont {S.~N.}\ \bibnamefont {Yannopoulos}}, \bibinfo {author}
  {\bibfnamefont {G.~A.}\ \bibnamefont {Voyiatzis}},\ and\ \bibinfo {author}
  {\bibfnamefont {T.}~\bibnamefont {Uruga}},\ }\href@noop {} {\bibfield
  {journal} {\bibinfo  {journal} {Phys. Rev. B}\ }\textbf {\bibinfo {volume}
  {82}},\ \bibinfo {pages} {155209} (\bibinfo {year} {2010})}\BibitemShut
  {NoStop}%
\bibitem [{\citenamefont {Momma}\ and\ \citenamefont
  {Izumi}(2011)}]{Momma:db5098}%
  \BibitemOpen
  \bibfield  {author} {\bibinfo {author} {\bibfnamefont {K.}~\bibnamefont
  {Momma}}\ and\ \bibinfo {author} {\bibfnamefont {F.}~\bibnamefont {Izumi}},\
  }\href@noop {} {\bibfield  {journal} {\bibinfo  {journal} {Journal of Applied
  Crystallography}\ }\textbf {\bibinfo {volume} {44}},\ \bibinfo {pages} {1272}
  (\bibinfo {year} {2011})}\BibitemShut {NoStop}%
\bibitem [{\citenamefont {Hu}\ \emph {et~al.}(2015)\citenamefont {Hu},
  \citenamefont {Vanacore}, \citenamefont {Yang}, \citenamefont {Miao},\ and\
  \citenamefont {Zewail}}]{jianbo2015}%
  \BibitemOpen
  \bibfield  {author} {\bibinfo {author} {\bibfnamefont {J.}~\bibnamefont
  {Hu}}, \bibinfo {author} {\bibfnamefont {G.~M.}\ \bibnamefont {Vanacore}},
  \bibinfo {author} {\bibfnamefont {Z.}~\bibnamefont {Yang}}, \bibinfo {author}
  {\bibfnamefont {X.}~\bibnamefont {Miao}},\ and\ \bibinfo {author}
  {\bibfnamefont {A.~H.}\ \bibnamefont {Zewail}},\ }\href@noop {} {\bibfield
  {journal} {\bibinfo  {journal} {ACS Nano}\ }\textbf {\bibinfo {volume} {9}},\
  \bibinfo {pages} {6728} (\bibinfo {year} {2015})}\BibitemShut {NoStop}%
\bibitem [{\citenamefont {Chen}\ \emph {et~al.}(2018)\citenamefont {Chen},
  \citenamefont {Li}, \citenamefont {Bang}, \citenamefont {Wang}, \citenamefont
  {Han}, \citenamefont {West}, \citenamefont {Zhang},\ and\ \citenamefont
  {Sun}}]{chen2018}%
  \BibitemOpen
  \bibfield  {author} {\bibinfo {author} {\bibfnamefont {N.-K.}\ \bibnamefont
  {Chen}}, \bibinfo {author} {\bibfnamefont {X.-B.}\ \bibnamefont {Li}},
  \bibinfo {author} {\bibfnamefont {J.}~\bibnamefont {Bang}}, \bibinfo {author}
  {\bibfnamefont {X.-P.}\ \bibnamefont {Wang}}, \bibinfo {author}
  {\bibfnamefont {D.}~\bibnamefont {Han}}, \bibinfo {author} {\bibfnamefont
  {D.}~\bibnamefont {West}}, \bibinfo {author} {\bibfnamefont {S.}~\bibnamefont
  {Zhang}},\ and\ \bibinfo {author} {\bibfnamefont {H.-B.}\ \bibnamefont
  {Sun}},\ }\href@noop {} {\bibfield  {journal} {\bibinfo  {journal} {Phys.
  Rev. Lett.}\ }\textbf {\bibinfo {volume} {120}},\ \bibinfo {pages} {185701}
  (\bibinfo {year} {2018})}\BibitemShut {NoStop}%
\bibitem [{\citenamefont {Hamann}(2013)}]{hamnn2013}%
  \BibitemOpen
  \bibfield  {author} {\bibinfo {author} {\bibfnamefont {D.~R.}\ \bibnamefont
  {Hamann}},\ }\href@noop {} {\bibfield  {journal} {\bibinfo  {journal} {Phys.
  Rev. B}\ }\textbf {\bibinfo {volume} {88}},\ \bibinfo {pages} {085117}
  (\bibinfo {year} {2013})}\BibitemShut {NoStop}%
\bibitem [{\citenamefont {Perdew}\ \emph {et~al.}(1997)\citenamefont {Perdew},
  \citenamefont {Burke},\ and\ \citenamefont {Ernzerhof}}]{perdew1997}%
  \BibitemOpen
  \bibfield  {author} {\bibinfo {author} {\bibfnamefont {J.~P.}\ \bibnamefont
  {Perdew}}, \bibinfo {author} {\bibfnamefont {K.}~\bibnamefont {Burke}},\ and\
  \bibinfo {author} {\bibfnamefont {M.}~\bibnamefont {Ernzerhof}},\ }\href@noop
  {} {\bibfield  {journal} {\bibinfo  {journal} {Phys. Rev. Lett.}\ }\textbf
  {\bibinfo {volume} {78}},\ \bibinfo {pages} {1396} (\bibinfo {year}
  {1997})}\BibitemShut {NoStop}%
\bibitem [{\citenamefont {Tangney}\ and\ \citenamefont
  {Fahy}(2002)}]{FahyPhysRevB.65.054302}%
  \BibitemOpen
  \bibfield  {author} {\bibinfo {author} {\bibfnamefont {P.}~\bibnamefont
  {Tangney}}\ and\ \bibinfo {author} {\bibfnamefont {S.}~\bibnamefont {Fahy}},\
  }\href@noop {} {\bibfield  {journal} {\bibinfo  {journal} {Phys. Rev. B}\
  }\textbf {\bibinfo {volume} {65}},\ \bibinfo {pages} {054302} (\bibinfo
  {year} {2002})}\BibitemShut {NoStop}%
\bibitem [{\citenamefont {Marini}\ and\ \citenamefont
  {Calandra}(2021)}]{marini2021_1}%
  \BibitemOpen
  \bibfield  {author} {\bibinfo {author} {\bibfnamefont {G.}~\bibnamefont
  {Marini}}\ and\ \bibinfo {author} {\bibfnamefont {M.}~\bibnamefont
  {Calandra}},\ }\href@noop {} {\bibfield  {journal} {\bibinfo  {journal}
  {Phys. Rev. B}\ }\textbf {\bibinfo {volume} {104}},\ \bibinfo {pages}
  {144103} (\bibinfo {year} {2021})}\BibitemShut {NoStop}%
\bibitem [{\citenamefont {Monacelli}\ \emph {et~al.}(2021)\citenamefont
  {Monacelli}, \citenamefont {Bianco}, \citenamefont {Cherubini}, \citenamefont
  {Calandra}, \citenamefont {Errea},\ and\ \citenamefont
  {Mauri}}]{monacelli2021}%
  \BibitemOpen
  \bibfield  {author} {\bibinfo {author} {\bibfnamefont {L.}~\bibnamefont
  {Monacelli}}, \bibinfo {author} {\bibfnamefont {R.}~\bibnamefont {Bianco}},
  \bibinfo {author} {\bibfnamefont {M.}~\bibnamefont {Cherubini}}, \bibinfo
  {author} {\bibfnamefont {M.}~\bibnamefont {Calandra}}, \bibinfo {author}
  {\bibfnamefont {I.}~\bibnamefont {Errea}},\ and\ \bibinfo {author}
  {\bibfnamefont {F.}~\bibnamefont {Mauri}},\ }\href@noop {} {\bibfield
  {journal} {\bibinfo  {journal} {Journal of Physics: Condensed Matter}\
  }\textbf {\bibinfo {volume} {33}},\ \bibinfo {pages} {363001} (\bibinfo
  {year} {2021})}\BibitemShut {NoStop}%
\bibitem [{\citenamefont {Baroni}\ \emph {et~al.}(2001)\citenamefont {Baroni},
  \citenamefont {de~Gironcoli}, \citenamefont {Dal~Corso},\ and\ \citenamefont
  {Giannozzi}}]{baroni2001}%
  \BibitemOpen
  \bibfield  {author} {\bibinfo {author} {\bibfnamefont {S.}~\bibnamefont
  {Baroni}}, \bibinfo {author} {\bibfnamefont {S.}~\bibnamefont
  {de~Gironcoli}}, \bibinfo {author} {\bibfnamefont {A.}~\bibnamefont
  {Dal~Corso}},\ and\ \bibinfo {author} {\bibfnamefont {P.}~\bibnamefont
  {Giannozzi}},\ }\href@noop {} {\bibfield  {journal} {\bibinfo  {journal}
  {Rev. Mod. Phys.}\ }\textbf {\bibinfo {volume} {73}},\ \bibinfo {pages} {515}
  (\bibinfo {year} {2001})}\BibitemShut {NoStop}%
\bibitem [{\citenamefont {Giannozzi}\ \emph {et~al.}(2009)\citenamefont
  {Giannozzi}, \citenamefont {Baroni}, \citenamefont {Bonini}, \citenamefont
  {Calandra}, \citenamefont {Car}, \citenamefont {Cavazzoni}, \citenamefont
  {Ceresoli}, \citenamefont {Chiarotti}, \citenamefont {Cococcioni},
  \citenamefont {Dabo}, \citenamefont {Corso}, \citenamefont {de~Gironcoli},
  \citenamefont {Fabris}, \citenamefont {Fratesi}, \citenamefont {Gebauer},
  \citenamefont {Gerstmann}, \citenamefont {Gougoussis}, \citenamefont
  {Kokalj}, \citenamefont {Lazzeri}, \citenamefont {Martin-Samos},
  \citenamefont {Marzari}, \citenamefont {Mauri}, \citenamefont {Mazzarello},
  \citenamefont {Paolini}, \citenamefont {Pasquarello}, \citenamefont
  {Paulatto}, \citenamefont {Sbraccia}, \citenamefont {Scandolo}, \citenamefont
  {Sclauzero}, \citenamefont {Seitsonen}, \citenamefont {Smogunov},
  \citenamefont {Umari},\ and\ \citenamefont {Wentzcovitch}}]{QE}%
  \BibitemOpen
  \bibfield  {author} {\bibinfo {author} {\bibfnamefont {P.}~\bibnamefont
  {Giannozzi}}, \bibinfo {author} {\bibfnamefont {S.}~\bibnamefont {Baroni}},
  \bibinfo {author} {\bibfnamefont {N.}~\bibnamefont {Bonini}}, \bibinfo
  {author} {\bibfnamefont {M.}~\bibnamefont {Calandra}}, \bibinfo {author}
  {\bibfnamefont {R.}~\bibnamefont {Car}}, \bibinfo {author} {\bibfnamefont
  {C.}~\bibnamefont {Cavazzoni}}, \bibinfo {author} {\bibfnamefont
  {D.}~\bibnamefont {Ceresoli}}, \bibinfo {author} {\bibfnamefont {G.~L.}\
  \bibnamefont {Chiarotti}}, \bibinfo {author} {\bibfnamefont {M.}~\bibnamefont
  {Cococcioni}}, \bibinfo {author} {\bibfnamefont {I.}~\bibnamefont {Dabo}},
  \bibinfo {author} {\bibfnamefont {A.~D.}\ \bibnamefont {Corso}}, \bibinfo
  {author} {\bibfnamefont {S.}~\bibnamefont {de~Gironcoli}}, \bibinfo {author}
  {\bibfnamefont {S.}~\bibnamefont {Fabris}}, \bibinfo {author} {\bibfnamefont
  {G.}~\bibnamefont {Fratesi}}, \bibinfo {author} {\bibfnamefont
  {R.}~\bibnamefont {Gebauer}}, \bibinfo {author} {\bibfnamefont
  {U.}~\bibnamefont {Gerstmann}}, \bibinfo {author} {\bibfnamefont
  {C.}~\bibnamefont {Gougoussis}}, \bibinfo {author} {\bibfnamefont
  {A.}~\bibnamefont {Kokalj}}, \bibinfo {author} {\bibfnamefont
  {M.}~\bibnamefont {Lazzeri}}, \bibinfo {author} {\bibfnamefont
  {L.}~\bibnamefont {Martin-Samos}}, \bibinfo {author} {\bibfnamefont
  {N.}~\bibnamefont {Marzari}}, \bibinfo {author} {\bibfnamefont
  {F.}~\bibnamefont {Mauri}}, \bibinfo {author} {\bibfnamefont
  {R.}~\bibnamefont {Mazzarello}}, \bibinfo {author} {\bibfnamefont
  {S.}~\bibnamefont {Paolini}}, \bibinfo {author} {\bibfnamefont
  {A.}~\bibnamefont {Pasquarello}}, \bibinfo {author} {\bibfnamefont
  {L.}~\bibnamefont {Paulatto}}, \bibinfo {author} {\bibfnamefont
  {C.}~\bibnamefont {Sbraccia}}, \bibinfo {author} {\bibfnamefont
  {S.}~\bibnamefont {Scandolo}}, \bibinfo {author} {\bibfnamefont
  {G.}~\bibnamefont {Sclauzero}}, \bibinfo {author} {\bibfnamefont {A.~P.}\
  \bibnamefont {Seitsonen}}, \bibinfo {author} {\bibfnamefont {A.}~\bibnamefont
  {Smogunov}}, \bibinfo {author} {\bibfnamefont {P.}~\bibnamefont {Umari}},\
  and\ \bibinfo {author} {\bibfnamefont {R.~M.}\ \bibnamefont {Wentzcovitch}},\
  }\href@noop {} {\bibfield  {journal} {\bibinfo  {journal} {Journal of
  Physics: Condensed Matter}\ }\textbf {\bibinfo {volume} {21}},\ \bibinfo
  {pages} {395502} (\bibinfo {year} {2009})}\BibitemShut {NoStop}%
\bibitem [{\citenamefont {Giannozzi}\ \emph {et~al.}(2020)\citenamefont
  {Giannozzi}, \citenamefont {Baseggio}, \citenamefont {Bonfà}, \citenamefont
  {Brunato}, \citenamefont {Car}, \citenamefont {Carnimeo}, \citenamefont
  {Cavazzoni}, \citenamefont {de~Gironcoli}, \citenamefont {Delugas},
  \citenamefont {Ferrari~Ruffino}, \citenamefont {Ferretti}, \citenamefont
  {Marzari}, \citenamefont {Timrov}, \citenamefont {Urru},\ and\ \citenamefont
  {Baroni}}]{QE2}%
  \BibitemOpen
  \bibfield  {author} {\bibinfo {author} {\bibfnamefont {P.}~\bibnamefont
  {Giannozzi}}, \bibinfo {author} {\bibfnamefont {O.}~\bibnamefont {Baseggio}},
  \bibinfo {author} {\bibfnamefont {P.}~\bibnamefont {Bonfà}}, \bibinfo
  {author} {\bibfnamefont {D.}~\bibnamefont {Brunato}}, \bibinfo {author}
  {\bibfnamefont {R.}~\bibnamefont {Car}}, \bibinfo {author} {\bibfnamefont
  {I.}~\bibnamefont {Carnimeo}}, \bibinfo {author} {\bibfnamefont
  {C.}~\bibnamefont {Cavazzoni}}, \bibinfo {author} {\bibfnamefont
  {S.}~\bibnamefont {de~Gironcoli}}, \bibinfo {author} {\bibfnamefont
  {P.}~\bibnamefont {Delugas}}, \bibinfo {author} {\bibfnamefont
  {F.}~\bibnamefont {Ferrari~Ruffino}}, \bibinfo {author} {\bibfnamefont
  {A.}~\bibnamefont {Ferretti}}, \bibinfo {author} {\bibfnamefont
  {N.}~\bibnamefont {Marzari}}, \bibinfo {author} {\bibfnamefont
  {I.}~\bibnamefont {Timrov}}, \bibinfo {author} {\bibfnamefont
  {A.}~\bibnamefont {Urru}},\ and\ \bibinfo {author} {\bibfnamefont
  {S.}~\bibnamefont {Baroni}},\ }\href@noop {} {\bibfield  {journal} {\bibinfo
  {journal} {The Journal of Chemical Physics}\ }\textbf {\bibinfo {volume}
  {152}},\ \bibinfo {pages} {154105} (\bibinfo {year} {2020})}\BibitemShut
  {NoStop}%
\bibitem [{\citenamefont {Edwards}\ \emph {et~al.}(2005)\citenamefont
  {Edwards}, \citenamefont {Pineda}, \citenamefont {Schultz}, \citenamefont
  {Martin}, \citenamefont {Thompson},\ and\ \citenamefont
  {Hjalmarson}}]{edwards2005}%
  \BibitemOpen
  \bibfield  {author} {\bibinfo {author} {\bibfnamefont {A.~H.}\ \bibnamefont
  {Edwards}}, \bibinfo {author} {\bibfnamefont {A.~C.}\ \bibnamefont {Pineda}},
  \bibinfo {author} {\bibfnamefont {P.~A.}\ \bibnamefont {Schultz}}, \bibinfo
  {author} {\bibfnamefont {M.~G.}\ \bibnamefont {Martin}}, \bibinfo {author}
  {\bibfnamefont {A.~P.}\ \bibnamefont {Thompson}},\ and\ \bibinfo {author}
  {\bibfnamefont {H.~P.}\ \bibnamefont {Hjalmarson}},\ }\href@noop {}
  {\bibfield  {journal} {\bibinfo  {journal} {Journal of Physics: Condensed
  Matter}\ }\textbf {\bibinfo {volume} {17}},\ \bibinfo {pages} {L329}
  (\bibinfo {year} {2005})}\BibitemShut {NoStop}%
\bibitem [{\citenamefont {Bianco}\ \emph {et~al.}(2017)\citenamefont {Bianco},
  \citenamefont {Errea}, \citenamefont {Paulatto}, \citenamefont {Calandra},\
  and\ \citenamefont {Mauri}}]{bianco2017second}%
  \BibitemOpen
  \bibfield  {author} {\bibinfo {author} {\bibfnamefont {R.}~\bibnamefont
  {Bianco}}, \bibinfo {author} {\bibfnamefont {I.}~\bibnamefont {Errea}},
  \bibinfo {author} {\bibfnamefont {L.}~\bibnamefont {Paulatto}}, \bibinfo
  {author} {\bibfnamefont {M.}~\bibnamefont {Calandra}},\ and\ \bibinfo
  {author} {\bibfnamefont {F.}~\bibnamefont {Mauri}},\ }\href@noop {}
  {\bibfield  {journal} {\bibinfo  {journal} {Physical Review B}\ }\textbf
  {\bibinfo {volume} {96}},\ \bibinfo {pages} {014111} (\bibinfo {year}
  {2017})}\BibitemShut {NoStop}%
\bibitem [{\citenamefont {Lewis}(1973)}]{lewis1973}%
  \BibitemOpen
  \bibfield  {author} {\bibinfo {author} {\bibfnamefont {J.~E.}\ \bibnamefont
  {Lewis}},\ }\href@noop {} {\bibfield  {journal} {\bibinfo  {journal} {physica
  status solidi (b)}\ }\textbf {\bibinfo {volume} {59}},\ \bibinfo {pages}
  {367} (\bibinfo {year} {1973})}\BibitemShut {NoStop}%
\bibitem [{\citenamefont {Wdowik}\ \emph {et~al.}(2014)\citenamefont {Wdowik},
  \citenamefont {Parlinski}, \citenamefont {Rols},\ and\ \citenamefont
  {Chatterji}}]{wdowik2014}%
  \BibitemOpen
  \bibfield  {author} {\bibinfo {author} {\bibfnamefont {U.~D.}\ \bibnamefont
  {Wdowik}}, \bibinfo {author} {\bibfnamefont {K.}~\bibnamefont {Parlinski}},
  \bibinfo {author} {\bibfnamefont {S.}~\bibnamefont {Rols}},\ and\ \bibinfo
  {author} {\bibfnamefont {T.}~\bibnamefont {Chatterji}},\ }\href@noop {}
  {\bibfield  {journal} {\bibinfo  {journal} {Phys. Rev. B}\ }\textbf {\bibinfo
  {volume} {89}},\ \bibinfo {pages} {224306} (\bibinfo {year}
  {2014})}\BibitemShut {NoStop}%
\bibitem [{\citenamefont {Shaltaf}\ \emph {et~al.}(2008)\citenamefont
  {Shaltaf}, \citenamefont {Durgun}, \citenamefont {Raty}, \citenamefont
  {Ghosez},\ and\ \citenamefont {Gonze}}]{shltaf2008}%
  \BibitemOpen
  \bibfield  {author} {\bibinfo {author} {\bibfnamefont {R.}~\bibnamefont
  {Shaltaf}}, \bibinfo {author} {\bibfnamefont {E.}~\bibnamefont {Durgun}},
  \bibinfo {author} {\bibfnamefont {J.-Y.}\ \bibnamefont {Raty}}, \bibinfo
  {author} {\bibfnamefont {P.}~\bibnamefont {Ghosez}},\ and\ \bibinfo {author}
  {\bibfnamefont {X.}~\bibnamefont {Gonze}},\ }\href@noop {} {\bibfield
  {journal} {\bibinfo  {journal} {Phys. Rev. B}\ }\textbf {\bibinfo {volume}
  {78}},\ \bibinfo {pages} {205203} (\bibinfo {year} {2008})}\BibitemShut
  {NoStop}%
\bibitem [{\citenamefont {Campi}\ \emph {et~al.}(2017)\citenamefont {Campi},
  \citenamefont {Paulatto}, \citenamefont {Fugallo}, \citenamefont {Mauri},\
  and\ \citenamefont {Bernasconi}}]{CampiPhysRevB.95.024311}%
  \BibitemOpen
  \bibfield  {author} {\bibinfo {author} {\bibfnamefont {D.}~\bibnamefont
  {Campi}}, \bibinfo {author} {\bibfnamefont {L.}~\bibnamefont {Paulatto}},
  \bibinfo {author} {\bibfnamefont {G.}~\bibnamefont {Fugallo}}, \bibinfo
  {author} {\bibfnamefont {F.}~\bibnamefont {Mauri}},\ and\ \bibinfo {author}
  {\bibfnamefont {M.}~\bibnamefont {Bernasconi}},\ }\href
  {https://doi.org/10.1103/PhysRevB.95.024311} {\bibfield  {journal} {\bibinfo
  {journal} {Phys. Rev. B}\ }\textbf {\bibinfo {volume} {95}},\ \bibinfo
  {pages} {024311} (\bibinfo {year} {2017})}\BibitemShut {NoStop}%
\bibitem [{\citenamefont {Xia}\ and\ \citenamefont {Chan}(2018)}]{xia2018}%
  \BibitemOpen
  \bibfield  {author} {\bibinfo {author} {\bibfnamefont {Y.}~\bibnamefont
  {Xia}}\ and\ \bibinfo {author} {\bibfnamefont {M.~K.~Y.}\ \bibnamefont
  {Chan}},\ }\href@noop {} {\bibfield  {journal} {\bibinfo  {journal} {Applied
  Physics Letters}\ }\textbf {\bibinfo {volume} {113}} (\bibinfo {year}
  {2018})}\BibitemShut {NoStop}%
\bibitem [{\citenamefont {Wang}\ \emph {et~al.}(2021)\citenamefont {Wang},
  \citenamefont {Wu}, \citenamefont {Zeng}, \citenamefont {Embs}, \citenamefont
  {Pei}, \citenamefont {Ma},\ and\ \citenamefont {Chen}}]{wang2021}%
  \BibitemOpen
  \bibfield  {author} {\bibinfo {author} {\bibfnamefont {C.}~\bibnamefont
  {Wang}}, \bibinfo {author} {\bibfnamefont {J.}~\bibnamefont {Wu}}, \bibinfo
  {author} {\bibfnamefont {Z.}~\bibnamefont {Zeng}}, \bibinfo {author}
  {\bibfnamefont {J.}~\bibnamefont {Embs}}, \bibinfo {author} {\bibfnamefont
  {Y.}~\bibnamefont {Pei}}, \bibinfo {author} {\bibfnamefont {J.}~\bibnamefont
  {Ma}},\ and\ \bibinfo {author} {\bibfnamefont {Y.}~\bibnamefont {Chen}},\
  }\href@noop {} {\bibfield  {journal} {\bibinfo  {journal} {npj Computational
  Materials}\ }\textbf {\bibinfo {volume} {7}},\ \bibinfo {pages} {118}
  (\bibinfo {year} {2021})}\BibitemShut {NoStop}%
\bibitem [{\citenamefont {Wuttig}\ and\ \citenamefont
  {Yamada}(2007)}]{wuttig2007}%
  \BibitemOpen
  \bibfield  {author} {\bibinfo {author} {\bibfnamefont {M.}~\bibnamefont
  {Wuttig}}\ and\ \bibinfo {author} {\bibfnamefont {N.}~\bibnamefont
  {Yamada}},\ }\href@noop {} {\bibfield  {journal} {\bibinfo  {journal} {Nature
  Materials}\ }\textbf {\bibinfo {volume} {6}},\ \bibinfo {pages} {824}
  (\bibinfo {year} {2007})}\BibitemShut {NoStop}%
\bibitem [{\citenamefont {Singh}(2013)}]{Singh_Sci_Rep}%
  \BibitemOpen
  \bibfield  {author} {\bibinfo {author} {\bibfnamefont {D.~J.}\ \bibnamefont
  {Singh}},\ }\href@noop {} {\bibfield  {journal} {\bibinfo  {journal} {Journal
  of Applied Physics}\ }\textbf {\bibinfo {volume} {113}},\ \bibinfo {pages}
  {203101} (\bibinfo {year} {2013})}\BibitemShut {NoStop}%
\bibitem [{\citenamefont {Setyawan}\ and\ \citenamefont
  {Curtarolo}(2010)}]{Setyawan2010}%
  \BibitemOpen
  \bibfield  {author} {\bibinfo {author} {\bibfnamefont {W.}~\bibnamefont
  {Setyawan}}\ and\ \bibinfo {author} {\bibfnamefont {S.}~\bibnamefont
  {Curtarolo}},\ }\href@noop {} {\bibfield  {journal} {\bibinfo  {journal}
  {Computational Materials Science}\ }\textbf {\bibinfo {volume} {49}},\
  \bibinfo {pages} {299} (\bibinfo {year} {2010})}\BibitemShut {NoStop}%
\bibitem [{\citenamefont {Kokalj}(1999)}]{KOKALJ1999176}%
  \BibitemOpen
  \bibfield  {author} {\bibinfo {author} {\bibfnamefont {A.}~\bibnamefont
  {Kokalj}},\ }\href@noop {} {\bibfield  {journal} {\bibinfo  {journal}
  {Journal of Molecular Graphics and Modelling}\ }\textbf {\bibinfo {volume}
  {17}},\ \bibinfo {pages} {176} (\bibinfo {year} {1999})}\BibitemShut
  {NoStop}%
\bibitem [{\citenamefont {Hase}\ \emph {et~al.}(2015)\citenamefont {Hase},
  \citenamefont {Fons}, \citenamefont {Mitrofanov}, \citenamefont {Kolobov},\
  and\ \citenamefont {Tominaga}}]{Hase2015}%
  \BibitemOpen
  \bibfield  {author} {\bibinfo {author} {\bibfnamefont {M.}~\bibnamefont
  {Hase}}, \bibinfo {author} {\bibfnamefont {P.}~\bibnamefont {Fons}}, \bibinfo
  {author} {\bibfnamefont {K.}~\bibnamefont {Mitrofanov}}, \bibinfo {author}
  {\bibfnamefont {A.~V.}\ \bibnamefont {Kolobov}},\ and\ \bibinfo {author}
  {\bibfnamefont {J.}~\bibnamefont {Tominaga}},\ }\href@noop {} {\bibfield
  {journal} {\bibinfo  {journal} {Nature Communications}\ }\textbf {\bibinfo
  {volume} {6}},\ \bibinfo {pages} {8367} (\bibinfo {year} {2015})}\BibitemShut
  {NoStop}%
\end{thebibliography}
\end{document}